\documentclass[12pt]{article}
\usepackage[dvips]{graphicx}

\begin{document}

\begin{center}
{\Large{\bf Test of \boldmath $\phi$ renormalization in nuclei through \\

\vspace{0.3cm}

\boldmath $\phi$ photoproduction }}

\vspace{0.3cm}

\end{center}

\vspace{1cm}

\begin{center}
{\large{E. Oset$^{1}$, M.J. Vicente Vacas$^{1}$, H.Toki$^{2}$
 , A. Ramos$^{3}$}}
\end{center}

\begin{center}
{\small{$^{1}$ \it Departamento de F\'{\i}sica Te\'orica and IFIC, \\
Centro Mixto Universidad de Valencia-CSIC, \\
Ap. Correos 22085, E-46071 Valencia, Spain}}

{\small{$^{2}$ \it Research Center for Nuclear Physics,
Osaka University, Ibaraki,\\
Osaka 567-0047, Japan}}

{\small{$^{3}$ \it Departament d'Estructura i Constituents de la Mat\`eria, \\
Universitat de Barcelona, \\
 Diagonal 647, E-08028 Barcelona, Spain}}
\end{center}

\vspace{1cm}

\begin{abstract} {\small{ We propose an experimental procedure to find out the
medium modifications of the $\phi$ meson. The reaction is inclusive $\phi$ 
photoproduction in nuclei, looking for $K^+ K^-$ pairs from the $\phi$ decay 
with total momentum smaller than 100-150 $MeV/c$, which are made possible at 
energies  of present laboratories from center of mass $\phi$ backward
production   and the help of Fermi motion. We have conducted a many body
calculation of the mass distribution of the $\phi$ adapted to the experimental 
set up of a recent JLAB experiment where the backwards $\phi$ photoproduction
has been measured. Using recent results for the in medium properties of the 
$\phi$, we find
that the width of the invariant mass distribution for photoproduction
on medium to heavy nuclei is larger than the free $\phi$ width by a
factor of two or more. }} 
\end{abstract}

\vspace{2cm}


\newpage

 The medium modification of vector mesons in nuclei is the subject of intense
study, particularly for the $\rho$ meson, both theoretical
\cite{beng,chanfray,johan,klinuc,mosel} and experimental \cite{hades,ceres}.  The
search for the medium modifications  proceeds experimentally mostly by means
of  back to back dilepton production in heavy ion reactions, which ensures the
$\rho$ production at rest and guarantees its decay inside the nucleus. Yet,
experiments are difficult to interpret since they require to include a
cocktail of background reactions out of which the $\rho$ contribution is only a
fraction of the total, and the present measurements are explained both in
terms of a drop of the $\rho$ mass \cite{rmass}, as well as  with just an
appreciable increase in the decay width \cite{johan}.

   The $\phi$ meson has also been the subject of some works
\cite{klinuc,Kuwabara:1995ms,Haglin:1995xu,Smith:1998xu,klilett,phinuc}
and the results of these papers point at a sizeable renormalization of 
the $\phi$ width and a small mass shift. Testing the $\phi$ properties
in  a nuclear medium is an interesting problem by itself, as well as for what
it bears in common with other vector mesons, which share some of  the
renormalization mechanisms of the $\phi$. The $\phi$ renormalization in matter
has also received attention in heavy ion collisions 
\cite{Lissauer:1991fr,Asakawa:1994nn,Chung:1997mp} 
since $\phi$ production has been advocated as one of the signatures for a 
possible transition to the quark gluon plasma.
 Another reason is the connection
with the changes of kaons inside matter, a subject of intense debate due to
strong deviations from the low density theorem already at very low
densities, as evidenced by the needed attraction to reproduce kaonic atom
data
\cite{gal,satoru,nieves,Gal:2001da}, as well as due to the possibility of having kaon
condensation in neutron stars at sufficiently large densities \cite{kaplan}.

 We propose here a simple and clean method for  
measuring the modifications of the $\phi$ meson in nuclei. The reaction is
inclusive $\phi$ photoproduction in nuclei looking only into the slow
$\phi$ mesons produced. The $\phi$ photoproduction on
proton targets is a subject of research in laboratories like Jefferson Lab
\cite{volker}, or  Spring8/Osaka\cite{nakano}, and extensive theoretical  studies have
been already performed \cite{titov}. Actually the experiment planned at 
Spring8/Osaka is done with CH2 targets, and hence data on protons as well
as in nuclei will be simultaneously accumulated.
    
   The reaction proposed is $\gamma A \to \phi X$ where the elementary process
is the $\gamma N \to \phi N$ reaction.  If we take the $\phi$ produced
backwards in the center of mass (CM) of the elementary reaction, the 
$\phi$ momentum in the
nucleon at rest frame (lab frame) is 235 $MeV/c$, 168 $MeV/c$ for photon
energies of 3.6 $GeV$ and  6 $GeV$,
respectively, and smaller at higher photon energies. Yet, it is possible to
have even smaller $\phi$ momenta in the nuclear reaction due to Fermi motion
if the production takes place on nucleons carrying a momentum opposite to the
one of the photon.  In this case, by putting a restriction on the momenta of
the outgoing $\phi$, we can guarantee that most of the $\phi$ produced would
decay inside the nucleus, thus showing the medium properties of the meson. We
shall see that by restricting the momentum to be smaller than 100-150 $MeV/c$ 
there is
still a  sizeable proportion of events to make the method of testing $\phi$
properties in a nuclear medium efficient from the
experimental point of view. The dependence of the cross section on the
invariant mass of the $\phi$ can be deduced in this case from its decay 
into $K^+ K^-$. The distortion of the low energy kaons, of about 
20 $MeV$ energy, can be easily handled because the quasielastic collisions 
are very much
hindered by Pauli blocking. On the other hand, the events in which the $K^-$ 
is absorbed are simply lost. This  results in a decrease of the cross 
section without affecting much the shape of the invariant mass 
distribution of the $\phi$ and it will be taken into consideration in the
calculations.

  In order to evaluate theoretically the inclusive nuclear photoproduction
cross section we apply many body techniques, successfully used in inclusive 
neutrino \cite{singh,kosmas} and photon \cite{rafa} nucleus reactions. We
start by evaluating the photon selfenergy in nuclear matter which accounts
for $\phi-ph$ production in intermediate states, see fig.1 or analogously the
$\gamma N \to \phi N$ reaction inside the nucleus.
In the evaluation of the photon selfenergy we will explicitly take into 
account the modification of the $\phi$ properties in the medium. For this
purpose we shall use the results for the mass (no change or negligible at
the low density involved in this reaction) and the width from refs. 
\cite{klilett,phinuc}. The width contains $\phi$ decay into $K,\bar K$
(modified in the medium) as  well as $\phi N \rightarrow K Y$ channels
($Y=\Lambda, \Sigma$ ).
We shall call $T$ the matrix element for the $\gamma N \to \phi N$  transition.
 Since
we are only interested in the slow $\phi$ coming from $\phi$ produced
backwards in the CM,  the dynamics of the process would enter through the 
elementary transition matrix element backwards which we call $C$.   We shall
also assume it to be the same for protons and neutrons for simplicity, but
even if it has different weights it would only show up in the total
strength of the cross section but not in the shape of the $\phi$ invariant
mass distribution, which is the one that will provide the medium properties
of the $\phi$ meson. 

 \begin{figure}[htb]
 \begin{center}
\includegraphics[height=9.cm,width=4.5cm,angle=0] {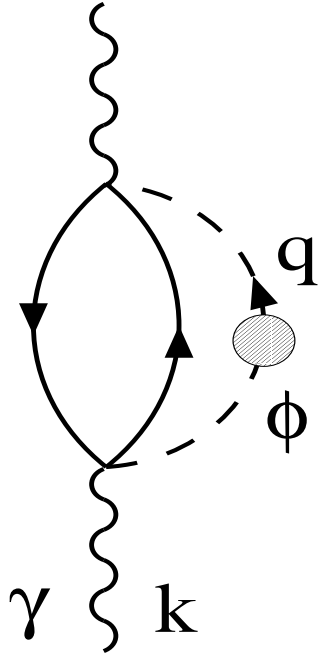}
 \caption{Photon selfenergy diagram accounting for a $\phi$ and a particle-hole
  in the intermediate states.}
 \label{fig:BSF}
 \end{center}
\end{figure}

       The photon selfenergy associated to the diagram of fig. 1 is given by
\begin{equation} -i\Pi(k,\rho)=\int \frac{d^4q}{(2\pi)^4} i U(k-q,\rho)
\frac{i(-iT)(-iT)}{q_0^2-\vec{q}\,^2-m_\phi^2+im_\phi\Gamma_\phi(\rho)}
\end{equation} where $U(p,\rho)$ is the relativistic Lindhard function for $ph$
excitation, for which we use the analytical expression given in \cite{singh}
(we multiply here the expression by a factor two, in order to account for
protons and  neutrons, and take the same value for the Fermi momentum of
protons and  neutrons, ($k_F=(3\pi^2\rho/2)^{1/3}$ ).

       Making use of Cutkosky rules to evaluate the imaginary part of $\Pi$
\begin{eqnarray} \Pi &\rightarrow& 2iIm\Pi\\\nonumber U(k-q)&\rightarrow&
2i\theta(k^0-q^0)Im U(k-q)\\\nonumber D(q) &\rightarrow& 2i\theta(q^0) Im D(q),
\end{eqnarray} where $D(q)$ is the $\phi$ propagator, we obtain

\begin{equation} Im \Pi(k,\rho)=\int \frac{d^4q}{(2\pi)^4} \theta(k^0-q^0)
\theta(q^0) Im U(k-q,\rho) \frac{2\mid T\mid^2 m_\phi\Gamma_\phi(\rho)}{\mid
q_0^2-\vec{q}\,^2-m_\phi^2+im_\phi\Gamma_\phi(\rho)\mid^2} \end{equation}

 Since $-Im \Pi/k $  gives the reaction probability per unit length for the
reaction  $\gamma N \to \phi N$   to occur inside the nucleus, the nuclear
cross section is then given by (see also \cite{rafa}):

\begin{equation} \sigma_A=-\int d^3r \frac{Im\Pi(k,\rho(\vec{r}))}{k}
\end{equation}

  One can see that in the limit of small densities, where the Lindhard function
takes the form

\begin{equation} Im U(k-q,\rho) \simeq -\frac{M}{E(\vec {k}-\vec {q})}\rho
\delta(k^0-q^0+M-E(\vec {k}-\vec {q})), \end{equation}
with $M,E(\vec{p})$ the mass and energy of the nucleon, respectively,
eq. (4) leads to the intuitive result $\sigma_A= A \sigma$, where $\sigma$ is
the elementary $\gamma N \to \phi N$ cross section . 

The differential cross section for $\phi$ photoproduction with the $\phi$
going forward in the nuclear lab frame, but with the small momenta
which correspond  to the backwards CM photoproduction, is given from eqs. (3),(4) 
by

\begin{eqnarray} \frac{d\sigma}{d\Omega} |_A &=& -\frac{1}{k}\int d^3 r
\frac{1}{(2\pi)^3} \int\frac{dq^0}{2\pi} \int q^2dq\theta(k^0-q^0) \theta(q^0) 
\\\nonumber &\times& Im U(k-q,\rho)  \frac{2\mid C\mid^2
m_\phi\Gamma_\phi(\rho)}{\mid
q_0^2-\vec{q}\,^2-m_\phi^2+im_\phi\Gamma_\phi(\rho)\mid^2} \end{eqnarray}

Even if we look at $\phi$ produced with small momentum, the $\phi$ will 
travel a certain distance inside the nucleus. The $\phi-ph $ production
occurs at the point $\vec{r}$ in the nucleus and hence the argument in the
Lindhard  function is $\rho(r)$. However, the $\phi$ is produced in the
point $\vec{r}$ but decays  at a point $\vec{r}+\vec{\Delta l}$ where 
$\vec{\Delta l}$ is the
average distance traveled by the $\phi$ in its life time in the nucleus.
Thus, for the width  of the $\phi$  (and eventually the density dependent mass)
appearing in the  $\phi$ propagator, we take the density at the average
point between the production and the decay, this is at $\vec{r}\,'=
\vec{r}+\vec{\Delta l/2}$, where 
$\vec{\Delta l}= \vec{q} /\omega(q) \Gamma(\rho)$,
with $\Gamma(\rho)$ the $\phi$ width at the production point and $\vec{q}$
taken forward in the nuclear lab frame. We can make some iteration in our
procedure in order to take this width also  at the intermediate point and our
final results are evaluated in that way.

      Taking now $\vec{b}$ as the impact parameter and replacing the
integration variable $dq$ by $ds$ 
($s=(q^0)^2-\vec{q}\,^2=M_I^2$), we obtain the final formula, which accounts
already for the cut in the $\phi$ momentum,

\begin{eqnarray} \frac{d\sigma}{d\Omega dM_I}  &=&
-\frac{2}{k(2\pi)^3}\int b d b \int dz  \int dq^0 \bar q M_I 
\theta(k^0-q^0)
\theta(q^0) \\\nonumber &\times& Im U(k-q,\rho(\vec r))\mid_{q=\bar q}
\frac{\bar q M_I \mid C\mid^2 m_\phi\Gamma_\phi(\rho(\vec{r}\,^\prime))
C_F C_A}{\mid
q_0^2-\vec{q}\,^2-m_\phi^2+im_\phi\Gamma_\phi(\rho(\vec{r}\,^\prime))\mid^2}
\end{eqnarray}
      with 
\begin{eqnarray} \bar q &=& (q_0^2-s)^{1/2}\\\nonumber
r&=&(b^2+z^2)^{1/2}\\\nonumber r'&=&(b^2+(z+\Delta l/2)^2)^{1/2}\\\nonumber
\Delta l&=&\frac{\bar q}{\omega(\bar q)}\frac{1}{\Gamma_\phi(\rho(r))}
\end{eqnarray} 
with the acceptance factor, $C_F =0$ for $q>q_{cut}$ and 1 otherwise. The factor
$C_A$ is the $K^-$ absorption factor. It is evaluated for each point $\vec{r}$
by assuming the $\phi$ to go forward. Then a random direction is taken for the
$K^-$ in the $\phi$ rest frame and the $K^-$ momentum is boosted to the $\phi$ 
moving frame. The surviving probability of this kaon is easily evaluated
knowing that the probability of absorption per unit length is
 $-Im \,\Pi_K$/ $p_K$, where $\Pi_K$ is the $K^-$ selfenergy which we take from
 \cite{angels}. Finally an average for all the original random momenta is taken.
  Another source of nuclear corrections is the Coulomb correction
 to the $K^+$ $K^-$ which will modify the invariant mass of the reconstructed
 $\phi$. An estimate can be done taking a couple of kaons going back to back
 from the center of the nucleus, assuming about 5 $MeV$ increase of energy to the
 $K^+$ and an equal decrease for the $K^-$ with the corresponding changes in the
 momenta.  The invariant mass change with respect to the free one is less than
 1 $MeV$. This magnitude is small compared to the strong interaction changes that
 we are discussing and hence we neglect it in the evaluations.
 
 The absorption of the $K^-$ is rather strong and only those $K^-$ produced
at the nuclear surface have a chance to survive. In our calculations the
average nuclear density for these observable events is around $\rho=\rho_0/4$.
According to recent selfconsistent calculations of the $K^-$ selfenergy
using the chiral Lagrangians \cite{angels,Schaffner-Bielich:2000cp,lutz} the   $K^-$ potential at this
density is of the order of -10 MeV. These results have been found 
consistent with $K^-$ atomic data \cite{satoru,nieves,Gal:2001da}. Hence,
the distortion  of the $K^-$ on their way out due to this potential is expected
to be small.

   We perform now calculations for a photon energy of 3.6 $GeV$, the energy of
the experiment performed by the CLAS collaboration at Jefferson Lab
\cite{volker}. The value of the constant $C$ can be
obtained from the data of this experiment. Taking $d \sigma / d t$ = 1
$nb/GeV^2$ at backward angles from \cite{volker} we obtain 
$\mid C \mid^2$=650 $nb$. One of the striking features of this
experiment is the relatively large backward cross section, which is two
orders of magnitude bigger  than what one would expect from the
extrapolation of the previous data at small angles \cite{previous}, or
theoretical expectations based on Pomeron exchange. On the other hand,
theoretical approaches based on two gluon exchange do predict such
enhancement of the backward cross section \cite{laget}. For us what matters
is that this cross section is large and measurable and, provided we find
nuclear cross sections of the same order of magnitude, the feasibility of
the experiment is guaranteed.

\begin{figure}[htb]
 \begin{center}
\includegraphics[width=13cm,angle=0] {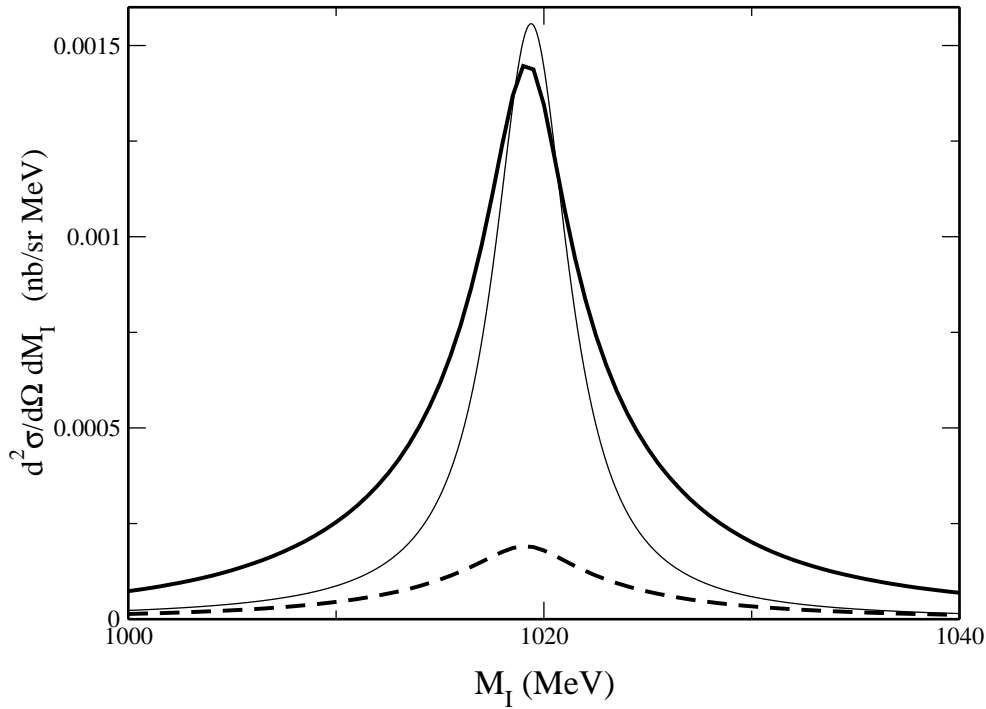}
 \caption{$\phi$ photoproduction cross section in $^{40}Ca$. The dashed line has
 a cut in the $\phi$ momentum of 100 $MeV/c$. The thick solid line is for a cut of
 150 $MeV/c$. The thin solid line is one third of the cross section calculated
 using the free $\phi$ width and a cut of 150 $MeV/c$. }
 \label{fig:BSF5}
 \end{center}
\end{figure}

\begin{figure}[htb]
 \begin{center}
\includegraphics[width=13cm,angle=0] {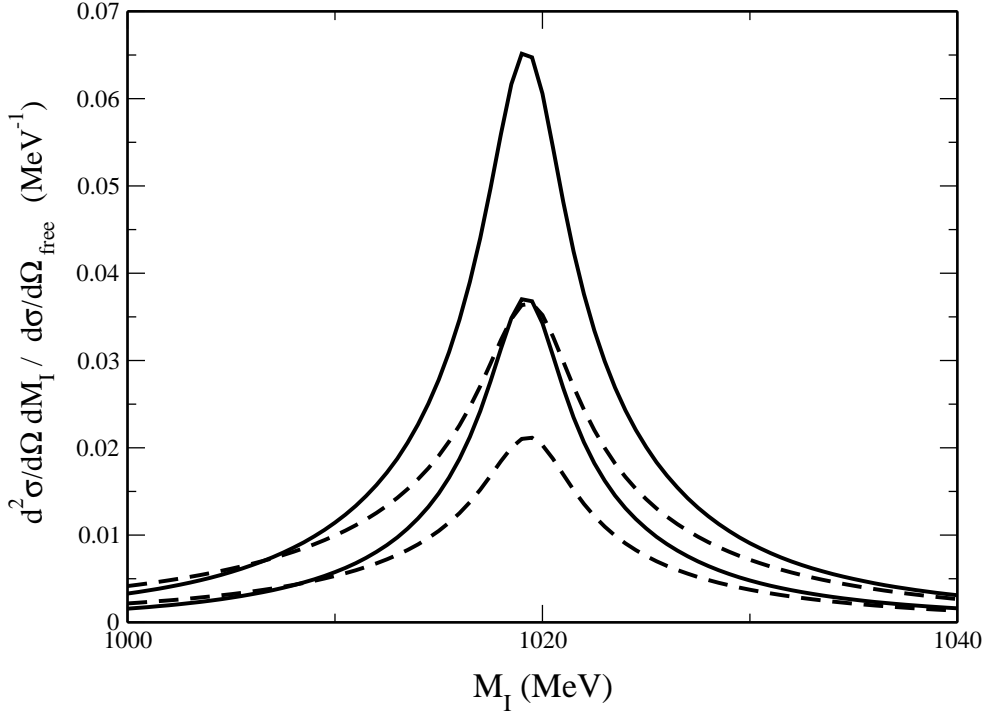}
 \caption{$\phi$ photoproduction cross section at $E_\gamma=3.6 \;GeV$ normalized 
 to the elementary one
 at backward angles. Solid lines:  $^{40}Ca$ (upper) and $^{16}O$ (lower).
 Dashed lines: Results  with the medium dependent width of \cite{klinuc,klilett}}
 \label{fig:BSF6}
 \end{center}
\end{figure}

   Our results for the photoproduction of the $\phi$ on $^{40} Ca$ are shown 
   in fig.2 at a photon energy $E_\gamma=3.6$ $GeV$ and two 
 $\phi$ momentum cuts of 100 $MeV/c$ and 150 $MeV/c$. The calculations have been 
 done using the density dependent $\phi$
width obtained  in \cite{phinuc} and a mass of the $\phi$ independent of the
density. The calculations of \cite{klinuc,klilett} indicate that the
variation of the mass is of the order of one percent. The
width of the $\phi$ obtained in \cite{phinuc} at $\rho=\rho_0$ is of the
order of 22 $MeV$, and the density dependence is not exactly linear. We shall also
show some cross sections (in fig. 3) using the results of \cite{klinuc,klilett} 
where the 
$\phi$ width is about 45 $MeV$ at $\rho=\rho_0$. For comparison  we
also show in fig. 2 the results obtained with the free width of the $\phi$. 
 We can see
that, for a same momentum cut, the use of the width of the  $\phi$ in
the medium reduces the cross section in about a factor three and leads to a
wider mass distribution.

The results of fig. 2 show an invariant mass distribution with a $\phi$
width  of about 10 $MeV$ for the momentum cut of 100 $MeV/c$.  Increasing the 
momentum cut  at 150 $MeV/c$ reduces
the width to about 8 $MeV$, nearly twice the free width, but it has
the advantage to increase the cross section by about a factor five, which is
desirable from the experimental point of view. This also means that moderate
increases in the cut can be done, at the expense of losing some width but 
increasing further the strength of the cross section.

It is worth mentioning that, should we have not included the cut in the
momentum of the $\phi$, the amount of events  where the $\phi$ decays outside
the nucleus, even concentrating on those coming from backward $\phi$
production in the CM frame, is quite large and the mass  distribution has a
width very close to the free one and is rather insensitive to the medium
corrections. Thus, the cut in the $\phi$ momentum is an essential ingredient
for the observation of the $\phi$ properties in the medium. A smaller 
momentum cut
further stresses the medium corrections at the cost of having fewer events,
but smaller cuts than 100 $MeV/c$ are still possible with enough statistics. 
With
higher energies of the photon beam one has more events with low $\phi$ momentum
but the backward photoproduction cross section might also decrease.  Knowledge
of these cross sections in a range of energies in the $GeV$ region would help 
in finding  an optimum photon energy for the present reaction.

In order to have a feeling about the feasibility of the experiment we
present the results in a different way, dividing the nuclear cross section
by the elementary  $d\sigma/d\Omega$  cross section in the lab
frame at backward CM angles, which we induce  easily from the experimental 
value of $d\sigma/dt$ of ref. \cite{volker}. 
The results corresponding to fig. 2 (with the momentum cut of 150 $MeV/c$)
are shown in fig. 3 in units of $MeV^{-1}$ for two nuclei. This is a convenient
representation, since the area of the curve gives us the fraction of the
elementary backward cross section that we obtain in the nuclear case with
the proposed set up.  We observe that we obtain a fraction of about 45
percent of the free cross section in the case of $^{16}O$ and about 85 percent
for the case of $^{40}Ca$. In the same figure we also show the results obtained
using the $\phi$ width in the medium of \cite{klinuc,klilett}. We observe in
this case that the width of the $\phi$ distribution is further increased and the
strength of the cross section at the peak decreased but is still sizeable.
 The results shown in fig. 3 thus indicate that the experimental observation 
 of sizeable
changes in the nuclear width of the $\phi$ should require about the same time as the
measurements done in \cite{volker}. 

      In summary, the reaction proposed of inclusive $\phi$ photoproduction
in nuclei, including a cut of about 100-150 $MeV/c$ for the $\phi$ momentum, 
proves to be sensitive to the $\phi$ properties inside nuclear matter.
 The large enhancements in the $\phi$ width predicted by theoretical
models, and the relatively large cross sections calculated,  call
 for the implementation of the experiment. 
 Furthermore, the data obtained could serve as
a test of the approaches used in the renormalization of other vector mesons, 
like
the $\rho$, and also as a further consistency test of  theories about 
the modification of kaon properties in nuclei where there is still much debate
\cite{gal,angels,lutz,koch,wolfram,caro}.

\vspace{3cm}

\section {Acknowledgments}
We would like to acknowledge useful discussions with V. Burkert and D. Tedeschi.
This work is partly supported by DGICYT contracts no. PB 96-0753 and  PB98-1247
and by the EEC-TMR Program Contract no. ERBFMRX-CT98-0169.

\end{document}